\documentclass[pra, twocolumn]{revtex4-1}

\usepackage[english]{babel}
\usepackage[utf8]{inputenc}
\usepackage{graphicx}
\usepackage[intlimits, sumlimits, namelimits]{mathtools}
\usepackage{physics}
\usepackage{mathrsfs}
\usepackage{amsfonts}
\usepackage{xcolor}
\usepackage{graphicx}

\renewcommand{\vec}{\vb}
\renewcommand{\imath}{i}
\newcommand{\PT}{$\mathcal{PT}$}

\definecolor{darkblue}{rgb}{0., 0.0, 0.4}

\newcommand{\bluetext}[1]{#1}
\newcommand{\bluetextt}[1]{#1}

\begin{document}

\title{Retardation-induced exceptional point}

\author{Alexey~A.~Dmitriev${}^{1}$}
\email{alexey.dmitriev@metalab.ifmo.ru}
\author{Mikhail~V.~Rybin${}^{1,2}$}

\affiliation{$^1$Department of Physics and Engineering, ITMO University, St Petersburg 197101, Russia}
\affiliation{$^2$Ioffe Institute, St Petersburg 194021, Russia}

\date{\today}

\begin{abstract}
Exceptional points in an optical dimer of spheres, which have the same size and operate in the spectral region of the dipolar resonance, are considered. By choosing different materials of these spheres, we can offset the radiative loss and create a gain-loss contrast to achieve a parity--time (\PT)-symmetric dimer. In this case, an exceptional point corresponds to the point where the \PT\ symmetry is broken. At the same time, if we consider a symmetric dimer, where both spheres are made of the same material (which may have a purely real dielectric constant), exceptional points occurring due to the radiative loss non-Hermiticity can also be observed.
 We study the transition between the two regimes and demonstrate that the exceptional point emerges due to the retardative nature of the coupling between the spheres, which makes the equation for eigenfrequencies nonlinear and allows it to have nontrivial solutions even when there is no contrast between the spheres.
\end{abstract}

\maketitle

  \section{Introduction}

The physics of non-Hermitian systems emerging several decades ago has uncovered multiple counter-intuitive phenomena such as the existence of non-decaying states in systems with the parity-time (\PT) symmetry and coalescence of the eigenstates at the exceptional point (EP), where the Hamiltonian of the problem becomes defective, i.e. cannot be diagonalized~\cite{bender1998real, bender2007making}.
In quantum physics, exceptional points lead to a critical behaviour in many-body systems~\cite{hanai2020critical}, and are related to transport phenomena, such as the Kondo effect in $f$-electron materials~\cite{michishita2020relationship}. Also, an exactly solvable attractor has been demonstrated for an exceptional point in a two-level system~\cite{li2021attractor}.

Since electromagnetic problems are inherently non-Hermitian, \PT\ symmetry and exceptional point  singularities gained considerable attention in photonics during the last decade~\cite{elganainy2018non,oezdemir2019parity,feng2017non,miri2019exceptional}.
In photonic systems, the exceptional point is usually associated with \PT\ symmetry, as the point where this symmetry is broken~\cite{heiss2012physics}. However, recently this definition was shown to be a non-universal one~\cite{zhang2020pt}.
For example, a \PT-symmetric system has been demonstrated, where variation of the system parameters in a way that the eigenvalues cross through the exceptional point does not break the \PT\ symmetry~\cite{zhong2019crossing}.

\begin{figure}[t!]
  \centering
  \includegraphics{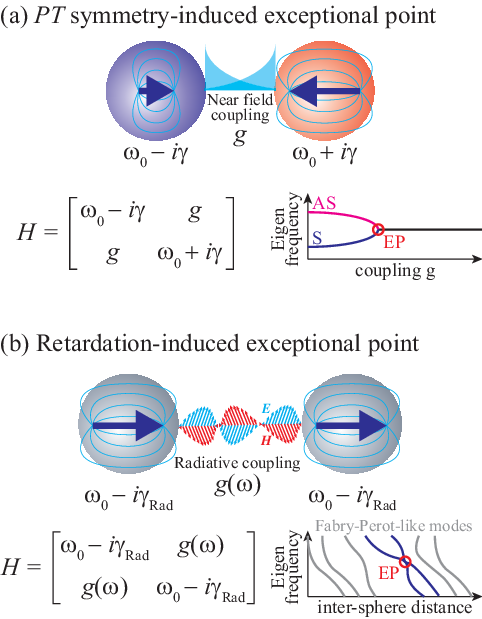}
  \caption{
    \label{fig:concept}
    Dimer eigenfrequency behavior depending on the type of optical coupling between the spheres. (a) For a \PT-symmetric dimer with compensated gain and loss, where the coupling is carried by near field, the Hamiltonian $H$ dictates a fork-like dependence of the eigenfrequencies, where, upon passing through the exceptional point (EP), the \PT\ symmetry is broken and eigenfrequency splitting disappears. (b) A dimer made of spheres made of the same lossless material and having the same radius, demonstrating an exceptional point due to the degeneration of the Fabry--Perot-like modes. Such EP exists due to the radiative coupling being retardative, i.e., frequency-dependent. However, a frequency-dependent dimer may be \PT-symmetric at the same time.
  }
\end{figure}

The reverse is also true. One does not need \PT\ symmetry for an exceptional point to exist. While it is well-known that systems with a loss and an even bigger loss behave similarly to \PT-symmetric systems (that are gain and loss), a system essentially needs some gain-loss contrast to be treated as \PT-symmetric~\cite{oezdemir2019parity}. Yet, multiple non-Hermitian systems without any gain-loss contrast have been shown to have exceptional points.
They include periodic arrays of cylinders~\cite{abdrabou2020exceptional}, insulator-metal-insulator plasmonic waveguides~\cite{min2020exceptional}, an array of four silicon disks~\cite{pichugin2021exceptional}, a non-symmetric dimer of dielectric cylinders with a real dielectric constant~\cite{valero2021exceptional}, a two-dimensional Fermi liquid confined in a slab~\cite{aquino2021probing}, random media with a real dielectric constant~\cite{bachelard2022coalescence}, a two-dimensional acoustic waveguide with arbitrary (such as both gain, or both lossy) admittance boundary conditions~\cite{perreydebain2022mode}, and a carefully designed three-way microstrip waveguide~\cite{yazdi2022sixth}. While in some systems the exceptional points emerge due to a nonlinear effect~\cite{fanisani2021level,li2021nonlinear}, in linear photonic systems without gain-loss contrast, the non-Hermiticity responsible for the emergence of the exceptional points is due to the radiative loss. In this sense, exceptional points occurring due to mode coupling inside a single dielectric particle with no material gain and loss, are even more illustrative. Such particles include a dumbbell-shaped resonator~\cite{yan2020shape}, a spheroid~\cite{bulgakov2021exceptional}, cylinders and conical particles~\cite{valero2022magnetoelectric}.

Therefore, exceptional points in photonic systems can exist without \PT\ symmetry and without any contrast at all. However, a simple two-level model prohibits this, as without any contrast the equation for the exceptional point will require zero coupling. Therefore, the underlying physics of the exceptional points in photonic systems without \PT\ symmetry remains unclear. There is also a question, whether the exceptional point in a \PT-symmetric system can be smoothly transformed to the exceptional point that can exist without contrast.


In this work, we demonstrate both the exceptional point associated with \PT\ symmetry breaking and an exceptional point occurring without a gain-loss contrast in the same system.
We consider a dimer of spheres \bluetext{of the same radius $R$ (see a scheme in Fig.~\ref{fig:concept}). Due to the scalability of the Maxwell equations, we choose $R$ as a characteristic scale and leave it as a variable throughout the paper}. In Sec.~\ref{sec:ham}, we construct a non-Hermitian effective Hamiltonian of the dimer using the dipole approximation and considering the longitudinally-coupled electric dipole modes, \bluetext{which we choose for simplicity. Note that exceptional points may exist in other dipole modes and in higher multipoles, however the corresponding effective Hamiltonians generally have larger dimensions. } In Sec.~\ref{sec:findep} we introduce a procedure for finding the exceptional point frequency of a dimer for arbitrary materials of spheres, treating the radiative losses and coupling rigorously. In Sec.~\ref{sec:pt} we demonstrate in detail the \PT\ symmetric regime, which is shown in Fig.~\ref{fig:concept}(a). It is achieved by introducing a gain-loss contrast as well as common material gain which compensates for the radiative loss. In Sec.~\ref{sec:rep} we analyze the case of a symmetric dimer, where both spheres are made of the same material, so no gain-loss contrast exists. We demonstrate that the exceptional point emerges due to the retardative nature of the coupling between the spheres, which makes the equation for eigenfrequencies nonlinear and allows it to have nontrivial solutions even when there is no contrast between the spheres. In conclusion, we demonstrate that there is no sharp transition that would clearly distinct between these two regimes, therefore the exceptional point in this dimer is tightly connected with the retardative coupling between the modes, which occurs through the surrounding continuum.

\section{Dimer effective Hamiltonian}
\label{sec:ham}
  We use the coupled dipole model described in details elsewhere~\cite{merchiers2007light}. By equating the incident field to zero, one obtains the following set of self-consistent equations that describe the eigenmodes:
 \begin{equation}
  \label{eq:coupleddipoles}
  \begin{aligned}
   \vec{p}_1 &= \alpha_{\text{E}1} \qty[\hat{G}(\vec{d})\vec{p}_2 - Z_0\vec{M}(\vec{d})\times\vec{m}_2], \\
   \vec{m}_1 &= \alpha_{\text{M}1} \qty[ \hat{G}(\vec{d})\vec{m}_2 + Y_0\vec{M}(\vec{d})\times\vec{p}_2], \\
   \vec{p}_2 &= \alpha_{\text{E}2} \qty[ \hat{G}(-\vec{d})\vec{p}_1 - Z_0\vec{M}(-\vec{d})\times\vec{m}_1], \\
   \vec{m}_2 &= \alpha_{\text{M}2} \qty[ \hat{G}(-\vec{d})\vec{m}_1 + Y_0\vec{M}(-\vec{d})\times\vec{p}_1],
  \end{aligned}
 \end{equation}
 where $\vec{p}_{1,2}$ are the electric dipole moments of the particles in the dimer, $\vec{m}_{1,2}$ are the magnetic dipole moments, $\vec{d}$ is the distance between the centers of the scatterers, $\alpha_{\text{E}1,2}$ are the electric polarizabilities, $\alpha_{\text{M}1,2}$ are the magnetic polarizabilities, 
 $Z_0 = \sqrt{\mu_0 / \epsilon_0}$, $Y_0 = Z_0^{-1}$, and $\hat{G}(\vec{r})$ and $\vec{M}(\vec{r})$ are the dyadic Green functions expressed as follows
  \begin{align*}
    \hat{G}(\vec{r}) &= k^2 \frac{e^{\imath k r}}{4\pi r} \qty[A(r)\hat{I} + B(r) \frac{\vec{r}\vec{r}^{\top}}{r^2}], \\
    \vec{M}(\vec{r}) &= k^2 \frac{e^{\imath kr}}{4\pi r} C(r) \frac{\vec{r}}{r}, \\
    A(r) &= 1 + \frac{\imath}{kr} - \frac{1}{k^2r^2}, \\
    B(r) &= -1 - \frac{3\imath}{kr} + \frac{3}{k^2r^2}, \\
    C(r) &= 1 + \frac{\imath}{kr},
  \end{align*}
  where $k = \omega/c$ and $r = \norm{r}$.

  Let us consider a dimer oriented along the $x$ axis, i.\,e., $\vec{d} = d\vec{e}_x$. By writing down the equations~\eqref{eq:coupleddipoles} explicitly for each coordinate, one can note that the $x$ components of the dipole moments do not couple to the $y$ and $z$ components. Also, the $x$ components of the electric dipole moments are decoupled from the magnetic dipoles and vice versa, hence electric longitudinal modes and magnetic longitudinal modes exist independently.
  Let us consider the electric longitudinal modes. For those, the coupled dipole equation describing the eigenmodes reads
  \begin{equation}
    \label{eq:coupleddipoleslong}
    \begin{bmatrix}
    1 & \alpha_{\text{E}1} \frac{e^{\imath kd}}{2\pi d^3} \qty(\imath kd - 1) \\
    \alpha_{\text{E}2} \frac{e^{\imath kd}}{2\pi d^3} \qty(\imath kd - 1) & 1
    \end{bmatrix}
    \begin{bmatrix}
      p_{1x} \\ p_{2x}
    \end{bmatrix}
    =
    0.
  \end{equation}
  To obtain an effective Hamiltonian describing the longitudinal electric eigenmodes of the dimer, we approximate the electric polarizabilities of the scatterers as simple poles\bluetext{, which are derived from the dipole Mie coefficients}:
  \begin{equation}
    \alpha_{\text{E}n} = \frac{3\imath}{2k^3} \frac{\gamma_n}{\omega - \omega_n},
  \end{equation}
  were $n = 1,2$ and $\gamma_n$ is the radiative decay, which we define as the decay of the dipole Mie resonance of a sphere with a real refractive index. 
\bluetext{
This approximation allows us to write analytical expressions for the exceptional point. However, it is not necessary for the exceptional points to exist. For increased precision, exact Mie coefficients may be used instead of simple poles.
}

\bluetext{
  The information about the radius and the refractive index of a sphere is therefore stored in the Mie coefficient.
  Finding a zero $\omega_1$ of the electric dipole Mie coefficient for a sphere with a refractive index $n = n' + \imath n''$ allows us to construct a mapping $n' + \imath n'' \mapsto \omega_1$ from the refractive index of an isolated sphere to its scattering pole $\omega_1$. We will later use this mapping to convert scattering poles determined from our model to the correspondent refractive indices $n_1$ and $n_2$ of the spheres.
}

  After multiplying the equation~\eqref{eq:coupleddipoleslong} by $\mathrm{diag}(\omega_1 - \omega, \omega_2 - \omega)$, the matrix takes the form $\hat{H} - \omega \hat{I}$, where $\hat{H}$ is the effective Hamiltonian
  \begin{equation}
    \label{eq:Hamiltonian}
    \hat{H} - \omega \hat{I} = \begin{bmatrix}
    \omega_1 -\omega & \gamma_1 \frac{3\imath e^{ikd}}{4\pi k^3 d^3} \qty(1 - \imath kd) \\
    \gamma_2 \frac{3\imath e^{ikd}}{4\pi k^3 d^3} \qty(1 - \imath kd) & \omega_2 - \omega
    \end{bmatrix}.
  \end{equation}

\section{Finding the exceptional point}
\begin{figure*}[t]
  \centering
  \includegraphics{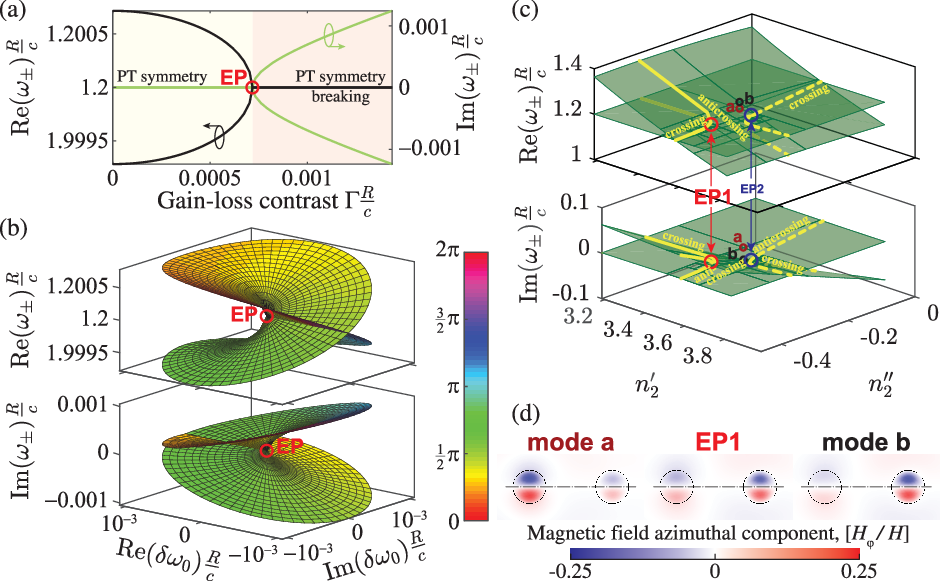}
  \caption{
    \label{fig:PTsym}
    Eigenfrequencies ($\omega_{\pm}$) of a \PT-symmetric dimer of spheres. (a) Real and imaginary parts of the eigenfrequencies of a \PT-symmetric dimer, as a dependence on a gain-loss contrast $\Gamma$. (b) Real and imaginary parts of the same dimer, as a dependence on the complex detuning $\delta\omega_0$ of the single sphere eigenfrequency from the exceptional point (EP). \bluetext {(c) Real and imaginary parts of the same dimer, as a dependence of the refractive index $n_2 = n_2' + \imath n_2''$ of the second sphere, calculated by the finite element method (FEM). \bluetextt{Dark-green thin solid lines show the eigenfrequencies obtained by FEM, while the filled rectangles show their bilinear interpolation.} The marks EP1,2 show two exceptional points. Thick solid and dashed yellow lines are guides to the eye indicating the crossings and anticrossings of the eigenfrequency sheets. (d) Mode field distributions corresponding to the frequencies marked by points a, b and EP1 in panel (c). The dash-dot line indicates the symmetry axis.}
  }
\end{figure*}

 \begin{figure*}
   \centering
   \includegraphics{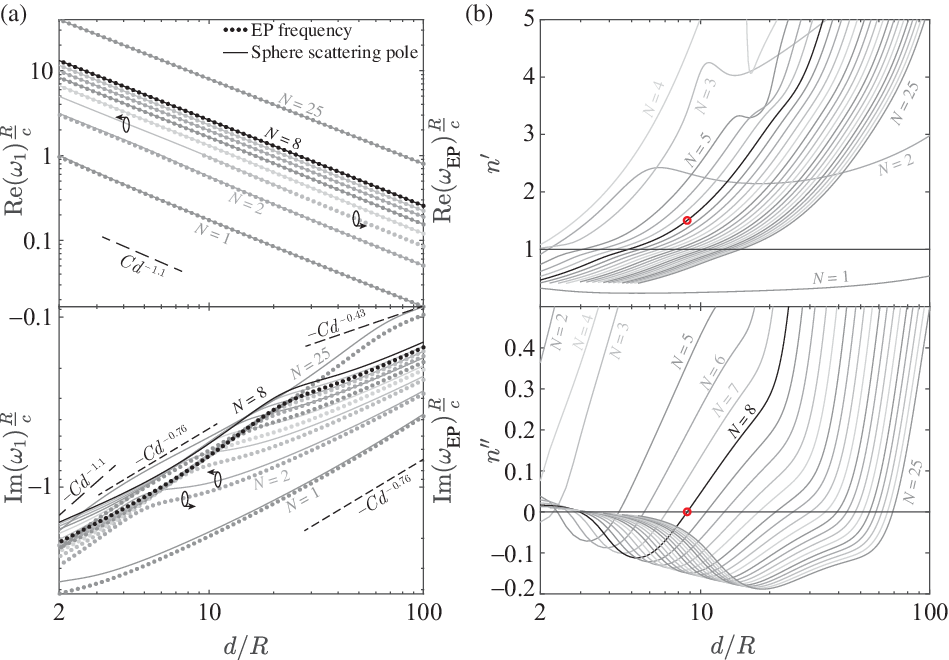}
   \caption{
     \label{fig:RetEP}
     Parameters of the retardation-induced exceptional points (REP) of symmetric dimers of spheres (radius $R$), separated by the center-to-center distance $d$.
     (a) REP eigenfrequencies ($\omega_{\text{EP}}$) corresponding to the coalescence of Fabry--Perot-like modes with different numbers ($N$), and the corresponding eigenfrequencies of the single sphere $\omega_{1}$, as a dependency on the center-to-center distance normalized by the radius.
     (b) Real and imaginary parts of the refractive index $n = n' + \imath n''$ of the sphere, corresponding to a scattering pole at the frequency $\omega_1$. The red circle marks a point of interest, corresponding to a dimer of silica spheres ($n = 1.55$).
   }
 \end{figure*}
\label{sec:findep}
    We are interested in finding the exceptional point, i.e., a point where two eigenfrequencies of the dimer coalesce into one. Consequently, this eigenfrequency $\omega_{\text{EP}}$ must be a double degenerate zero of the function
    \begin{equation}
    \label{eq:det}
      f(\omega) = \det(\omega \hat{I} - \hat{H}).
    \end{equation}
\bluetextt{
We note that, if exact Mie coefficients are used instead of simple poles, $f(\omega)$ has to be redefined as the determinant of the coupled multipole matrix.
}

    Using a Taylor expansion around $\omega_{\text{EP}}$, one may demonstrate that the zero will be double degenerate if 
    \begin{equation}
      \label{eq:ff}
      \left\lbrace
      \begin{array}{l}
      f(\omega_{\text{EP}}) = 0, \\
      f'(\omega_{\text{EP}}) = 0.
      \end{array}
      \right.
    \end{equation}
    We may use this system of equations to find the parameters corresponding to the exceptional point. Let us assume $\omega_{\text{EP}}$ and the distance $d$ between the scatterers as parameters, and the scattering poles $\omega_{1,2}$ of the isolated scatterers as unknowns.
  We also use the following approximation for the radiative decay:
  \begin{equation}
    \gamma_{1,2}(\omega_{1,2}) \simeq \gamma(\Re{\omega_{\text{EP}}}),
  \end{equation}
  which is justified by the fact that $\Re{\omega_{\text{EP}}} \simeq \Re{\omega_{1,2}}$ and that the radiative loss $\gamma$ is mainly determined by the real part of the scattering pole $\omega_{1,2}$. \bluetextt{Then the following approximation holds in our region of interest:}
  \begin{align*}
    \frac{\gamma_{1,2}R}{c} \simeq
  \exp\bigg[&-\ln\bigg(6.8\Re{\frac{\omega_{\text{EP}}R}{c}}^{-7.6}+\\
  +3.2\Re{\frac{\omega_{\text{EP}}R}{c}}^{-1.4}\bigg)
            &+ \sin\qty(1.2\ln\Re{\frac{\omega_{\text{EP}}R}{c}} - 2.5)\bigg],
  \end{align*}
  \bluetextt{see Supplementary Document for the verification~\cite{supplementary}.}
    
    By solving the system~\eqref{eq:ff} for $\omega_{1,2}$ we obtain the following expressions:
    \begin{equation}
      \label{eq:EPparam}
      \begin{aligned}
    \omega_{1,2} = \omega_{\text{EP}} + \frac{c}{d}u(\omega_{\text{EP}}) \frac{a(\omega_{\text{EP}})}{b(\omega_{\text{EP}})} \qty(1 \pm \sqrt{1 + \frac{1}{u(\omega_{\text{EP}})}}) \\
      \end{aligned}
    \end{equation}
    where
    \begin{align*}
      u(\omega) &= \qty[\frac{3}{4\pi}\frac{\sqrt{\gamma_1\gamma_2}}{c}d
     \left(\frac{\omega}{c}d\right)^{-3}
     b(\omega)
     \exp(i\frac{\omega}{c}d)]^2, \\
      a(\omega) &= 1 - \imath \frac{\omega}{c}d, \\
      b(\omega) &= \frac{\omega}{c}d - 3 \left(\frac{\omega}{c}d\right)^{-1} \left(1 - \imath \frac{\omega}{c}d\right).
    \end{align*}

  The equation~\ref{eq:EPparam} determines the scattering poles of two isolated spheres, which, being brought together to the distance $d$ between their centers, create an exceptional-point dimer with a single eigenmode at the frequency $\omega_{\text{EP}}$.
     \bluetext{At this point, the eigenfrequencies of the system coalesce to create a single mode, which is described by a degenerate eigenvector $\mathbf{v}$
     \begin{align*}
      \mathbf{v} &= \begin{bmatrix}
        1, & 
        -v(\omega_{\text{EP}}) + \sqrt{\qty(v(\omega_{\text{EP}}))^2 - 1}
        \end{bmatrix}^{\top},\\
      v(\omega) &= \frac{3}{4\pi} \frac{\sqrt{\gamma_1\gamma_2}}{c}d \frac{e^{\imath \frac{\omega}{c}d}}{\qty(\frac{\omega}{c}d)^4} \qty[\qty(\frac{\omega}{c}d)^2 + 3\frac{\omega}{c}d -3].
     \end{align*}
   }

 \begin{figure*}[p]
   \centering
   \includegraphics{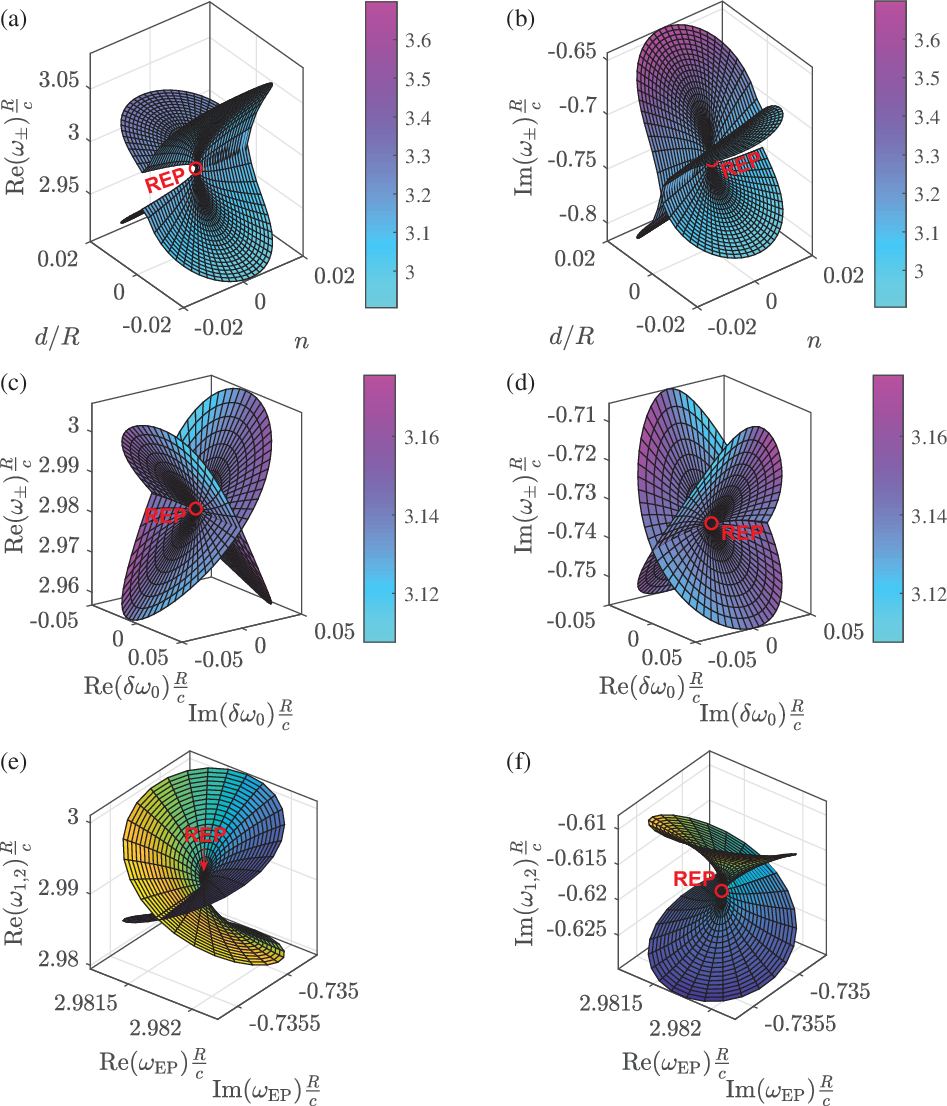}
   \caption{
     \label{fig:Whirls}
     Frequency sheet topologies around the retardation-induced exceptional point (REP) in a symmetric dimer of silica spheres ($n = 1.55$) of radius $R$, separated by a center-to-center distance $d = 8.757R$. (a,b) Real and imaginary parts of the eigenfrequencies $\omega_{\pm}$ of symmetric dimers with a detuned refractive index $n = 1.55 + \delta n$ and distance $d = 8.757R + \delta d$. (c,d) Real and imaginary parts of the eigenfrequencies, as a dependence of a complex detuning $\delta\omega_0$ introduced into the single sphere scattering pole. (e,f) Scattering poles $\omega_{1}$ and $\omega_{2}$ of spheres that, being brought together to a center-to-center distance $d = 8.757R$, create a dimer with an exceptional point at the frequency $\omega_{\text{EP}}$. }
 \end{figure*}
     
  \section{Parity-time symmetry}
  \label{sec:pt}
  Let us consider a limiting case of Eq.~\eqref{eq:EPparam} which corresponds to the parity-time symmetry breaking. In order to achieve a parity-time symmetry regime, one has to introduce gain into one of the scatterers and loss into another one. However, in the case of an open system, such as a dimer, a radiative loss exists in both of the particles, violating the parity-time symmetry of the system. Due to the retardation, the radiative loss depends on the materials of the particles and on the distance between them, so the problem of finding the frequencies in a \PT-symmetric dimer becomes self-consistent. In this case, we reformulate it as follows: let us search for the materials of the `gain' and `loss' scatterers, which, being brought at a distance $d$, will create a dimer with two eigenmodes at real frequencies, which are a feature of the \PT\ symmetry regime. Finally, these two eigenmodes will coalesce at an exceptional point.

  To demonstrate this behaviour in our system, we substitute the `target' exceptional point frequency $\omega_{\text{EP}} = 1.2c/R$ \bluetextt{(which corresponds to the vacuum wavelength 530~nm for $R = 100~\text{nm}$)} to Eq.~\eqref{eq:EPparam}. \bluetextt{In our previous work~\cite{dmitriev2023towards}, a \PT-symmetric exceptional point has been demonstrated for a perovskite with a refractive index equal to $2$ at this frequency.}  The conventional \PT-symmetric case assumes that, at the exceptional point, the scattering poles of the individual scatterers must be $\omega_{1,2} = \omega_{\text{EP}} \pm \imath g$, where $g$ is the coupling strength. However, in an open system, according to Eq.~\eqref{eq:EPparam}, \bluetext{off-diagonal elements of the Hamiltonian are generally complex values that change with the frequency and distance, so they cannot be treated as a real coupling constant. To demonstrate the conventional \PT-symmetric behaviour, we use the contrast between the poles at the exceptional point to define the coupling constant \bluetextt{$g = (\omega_1 - \omega_2)/2i$}. \bluetextt{To achieve a purely real coupling constant, we consider the distance $d = 5.101R$. A complex coupling constant and a complex exceptional point frequency cases are considered in the Supplementary Document~\cite{supplementary}, as well as the dependency of the complex coupling constant on the distance}.  \bluetextt{For the chosen exceptional point frequency, the radiative decays are equal to $\gamma_{1,2} = 0.1112c/R$.} According to Eq.~\eqref{eq:EPparam}, the scattering poles become $\omega_1 = (1.2 + 0.0007i)c/R$, which corresponds to $n_1 = 3.5829 - 0.3127i$, and $\omega_2 = (1.2-0.0007i)c/R$ that corresponds to $n_2 = 3.5826 - 0.3087i$. For the chosen distance, the coupling constant $g = 0.0007$ \bluetextt{indeed} turns out to be a purely real number.}

  We then substitute the obtained poles into the Hamiltonian~\eqref{eq:Hamiltonian} and search for its eigenvalues. First, we consider the scattering poles of the spheres as having a gain-loss contrast $\Gamma$ \bluetext{(which is a sweep parameter): $\omega_{1,2}' = \omega_{\text{EP}} \pm \imath \Gamma$,} and search for the dimer eigenfrequencies \bluetext{for each $\Gamma$}. The result is shown in Fig.~\ref{fig:PTsym}, and demonstrates a typical crossing-to-anticrossing transition, which is a characteristic sign of the \PT\ symmetry breaking. Then we consider a complex detuning of the scattering poles from the EP value: $\omega_{1,2}' = \omega_{1,2} \pm \delta\omega_0/2$. Fig.~\ref{fig:PTsym}(b) shows the dimer eigenfrequencies as a dependence on the complex detuning $\delta\omega_0$. The topology of the eigenfrequency sheets resembles that of the Riemann surface of the complex square root function, which is a characteristic feature of an EP.

  \bluetext{To verify our analytical model, we have compared its results with full-wave simulations in COMSOL Multiphysics. We used the exceptional point search algorithm that successively builds a denser grid in the parameter space around the EP~\cite{dmitriev2021finding}. The underlying model consisted of two spheres of the same radius $R$ with the distance $d = 5.101R$ between their centers. The refractive index $n_1 = 3.58 - 0.31i$ of the first sphere has been fixed, while for the second sphere it was varied. To select the longitudinal electric dipole mode and reduce the computational complexity, axial symmetry around the $x$ axis was utilized. The model was treated as axisymmetric with the azimuthal number $m = 0$, and the axial and radial components of the magnetic field vector as well as the electric field azimuthal component were taken as zero. 
  }

  \bluetext{
  Fig.~\ref{fig:PTsym}(c) shows the obtained eigenfrequency sheets. In the given search region, two exceptional points exist: 
  $\omega_{\text{EP}} = (1.20 - 0.01i)c/R$ at $n_2 = 3.60 - 0.252i$
  marked as EP1 in the figure, and
  $\omega_{\text{EP}} = (1.20 + 0.01i)c/R$ at $n_2 = 3.58 - 0.367i$
  marked as EP2. The exceptional point previously found by the analytical model is characterized by $n_1'' < n_2''$ and $n_1' \simeq n_2'$. Therefore, it corresponds to EP1. The obtained frequency and refractive index of the second sphere agree with the model with good precision, validating our approximations. \bluetextt{We note that both spheres need to be made of gain media to achieve the \PT~symmetry regime, to compensate for the radiative damping: $n_1 = 3.58-0.31i$, having a greater gain, corresponds to the resonator with gain, while $n_2 = 3.585 - 0.23i$ corresponds to the resonator with loss. We also note that the contrast in the imaginary parts of the spheres (which corresponds to the gain-loss contrast), is increased compared to the analytical model.}
  }

  \bluetext{
  Fig.~\ref{fig:PTsym}~(d) shows the mode field distributions at $n_2 = 3.585 - 0.23i$ (modes a and b in the figure) as well as the exceptional point mode at EP1. We note the similarity in the field distributions that indicate the degeneration of the eigenvector describing the mode at EP.
  }

  \section{Retardation-induced contrastless EP}
  \label{sec:rep}

  Another special case is observed when 
  \begin{equation}
    \label{eq:RetEPa}
    u(\omega_{\text{EP}}) = -1.
  \end{equation}
  At this case, the scattering poles $\omega_1$ and $\omega_2$ coincide:
  \begin{equation}
    \label{eq:RetEPb}
    \omega_{1,2} = \omega_{\text{EP}} + \frac{c}{d}u(\omega_{\text{EP}}) \frac{a(\omega_{\text{EP}})}{b(\omega_{\text{EP}})}.
  \end{equation}
  In other words, an exceptional point exists at a dimer made of two exactly equivalent spheres, without any refractive index contrast or gain-loss contrast. Such behavior has been observed in a dimer of infinite cylinders~\cite{dmitriev2021optical} and in a dimer of silicon nanorods~\cite{valero2021exceptional}. 

  To understand the nature of this kind of an exceptional point, let us consider a conventional two-level model with gain and loss~\cite{oezdemir2019parity}, but without any contrast:
  \begin{equation}
    \hat{H} - \omega \hat{I} = \begin{bmatrix}
      \omega_0 - \imath\gamma - \omega & g \\
      g & \omega_0 - \imath\gamma - \omega
    \end{bmatrix}.
  \end{equation}
  One can immediately see that an exceptional point cannot exist in such a model. Indeed, the eigenfrequencies are $\omega = \omega_0 - \imath\gamma \pm g$, and may only coalesce when the coupling $g$ is zero. However, in an electromagnetic system such as the dimer considered, the coupling $g$ depends on the frequency: $g = g(\omega)$, as may be seen from Eq.~\eqref{eq:Hamiltonian}. In this case, the equation determining the eigenfrequencies becomes nonlinear
  \begin{equation}
    \omega = \omega_0 - \imath\gamma \pm g(\omega),
  \end{equation}
  allowing the eigenfrequencies to coalescence even when $g \neq 0$. Therefore this kind of exceptional point is possible due to the retardation, so we will further call it a \emph{retardation-induced exceptional point} (REP).

  We use Eq.~\eqref{eq:RetEPa} to find the REP frequency $\omega_{\text{EP}}$ as a function of the distance $d$. By transforming it to the form
  \begin{equation}
    \label{eq:RetEPs}
    \imath\frac{\omega_{\text{EP}}}{c}d = \imath \pi N - \ln\qty(\frac{\imath\qty[\frac{\omega_{\text{EP}}}{c}d]^3}{b(\omega_{\text{EP}}) \frac{\sqrt{\gamma_1\gamma_2}}{c}d}) \quad (N \in \mathbb{Z}),
  \end{equation}
  one can show that it has an infinite number of solutions, each corresponding to the coalescence of a pair of Fabry--Perot-like modes with the number $N$. The solutions of Eq.~\eqref{eq:RetEPs} corresponding to different $N$ are shown with a dotted line in~Fig.~\ref{fig:RetEP}(a). The corresponding eigenfrequencies $\omega_{1,2}$ of the isolated scatterers 1 and 2, which are equivalent, are calculated using Eq.~(\ref{eq:RetEPb}) and shown with solid lines in Fig~\ref{fig:RetEP}(a).
  We note that $\Re{\omega_1} = \Re{\omega_2} = \Re{\omega_{\text{EP}}}$ (within $0.1\%$ accuracy), while $\Im{\omega_1} = \Im{\omega_2} \simeq K \Im{\omega_{\text{EP}}}$, where $1 < K < 1.6$.
  
  \bluetext{Using the mapping from a scattering pole to the refractive index}, we find the refractive index $n' + \imath n''$ of the spheres in the REP dimer. The results are shown in Fig.~\ref{fig:RetEP}(b). We note that the real part $n'$ of the refractive index has to be increased to achieve a REP in a dimer with a greater distance. As for the imaginary part, one may expect that a larger gain would be necessary at larger distances, however, strikingly, loss has to be introduced for larger distances.
  For most modes, there is also a point, where a purely real refractive index is required to achieve a REP. In this regard, the mode $N = 8$ is especially interesting, as the corresponding REP $\omega_{\text{EP}} = (2.982 - 0.7352\imath)c/R$ can be realized in a dimer of silica ($n = 1.55$) spheres, separated by the distance $d = 8.757R$. 
  
  Let us study the behavior of the REP frequency and the eigenfrequencies in the vicinity of the REP in more detail. First, we consider the dependence of the REP frequency on the dimer refractive index $n$ and the distance between spheres $d$ for symmetric dimers with a purely real refractive index of the sphere material. The resulting REP frequency sheets is shown in Figs.~\ref{fig:Whirls}(a,b). One can notice the resemblance of their topology to that of the usual EP eigenfrequency sheets. However, the eigenfrequency sheets themselves demonstrate a topology, which is shown in Fig.~\ref{fig:Whirls}, that is strikingly different from the conventional case. Instead of a crossing-to-anticrossing transition, an anticrossing-to-anticrossing transition is observed. This behaviour shows some similarity to higher-order exceptional points. Indeed, in this case the REP is achieved by matching two pairs of complex frequencies at once: the scattering poles have the same frequency ($\omega_1 = \omega_2$) and the exceptional point frequency itself is `at the exceptional point', but with the eigenfrequencies playing the role of the parameter space.
  Finally, we consider the dependency of the scattering pole frequencies $\omega_{1,2}$ on the exceptional point frequency $\omega_{\text{EP}}$, according to Eq.~\eqref{eq:EPparam}. 
  The resulting frequency sheets, shown in Figs.~\ref{fig:Whirls}(e,f), also demonstrate a square-root-like topology around the REP. Thus, the REP may be thought of as a point of double EP-like degeneration: there is an eigenfrequency coalescence in the parameter space, and, at the same time, there is a parameter coalescence at the eigenfrequency space.

  \begin{figure}[t]
    \includegraphics{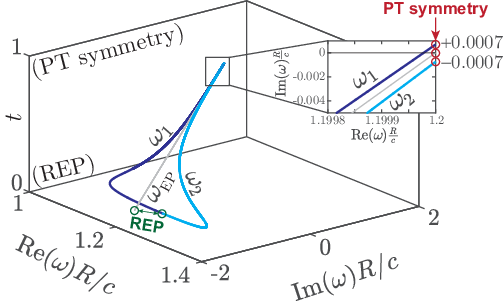}
    \caption{\label{fig:transition} \bluetextt{Transition between the \PT symmetry-induced exceptional point (marked ``PT symmetry'') where $\omega_1 = \omega_2^{*}$ and the retardation-induced exceptional point (marked ``REP''), where $\omega_1 = \omega_2$ for a dimer with a fixed distance between sphere centers $d = 5.101R$, where $R$ is the sphere radius. The exceptional point frequency $\omega_{\text{EP}}$ (shown with a thin line) is varied linearly, while the according scattering poles $\omega_1$ and $\omega_2$ (shown with thick lines) are tuned according to Eq.~\eqref{eq:EPparam}. The inset shows the vicinity of the \PT symmetry point upscaled.}}
  \end{figure}

\section{Discussion and conclusion}
  We have studied the exceptional points occurring in the dimers of spheres with gain and loss. Using the dipole approximation, we have derived the analytical formulae which connect the isolated sphere scattering poles, the distance between the centers of the spheres, and the exceptional point frequency. Two particular cases have been identified and compared: parity-time symmetry breaking and the so-called \emph{retardation-induced exceptional point}. The exceptional points existing at the systems with gain and loss at the parity-time symmetry breaking point are well-known, and exist at the point where the gain-loss contrast becomes equal to the coupling strength. On the contrary, the retardation-induced exceptional points exists in dimers with no contrast whatsoever. We have demonstrated a retardation-induced exceptional point in a dimer of identical spheres $R$ made of a glass with a refractive index $n = 1.55$ placed at a distance $d = 8.757R$ between their centers. In our previous work~\cite{dmitriev2021optical}, a retardation-induced exceptional point has been observed in a dimer of identical infinite cylinders of radius $R$ with a refractive index $n = 4.992$ placed at a distance $d = 36.44R$ between their axes. We have shown that such exceptional points appear because the coupling constant is a frequency-dependent value in retardative systems such as dimers.
  
  The two-level model with a frequency-dependent coupling describes both kinds of exceptional points, \bluetext{ and reveals a smooth transition exists between them. For example, the \PT-symmetric exceptional point observed for $d = 5.101R$ and $\omega_{1,2} = (1.2 \pm 0.0007i)c/R$, which was considered in Sec.~\ref{sec:pt}, can be smoothly continued by varying the exceptional point frequency in Eq.~\eqref{eq:EPparam} from $\omega_{\text{EP}} = 1.2c/R$ to $\omega_{\text{EP}} = (1.16 - 1.32i)c/R$, where it becomes a retardation-induced exceptional point. The corresponding poles coalesce at that point: $\omega_1 = \omega_2 = (1.19 - 1.09i)c/R$. \bluetextt{This transition is shown in Fig.~\ref{fig:transition} for a linear variation of $\omega_{\text{EP}}$ between the \PT-symmetric and retardation-induced exceptional points, however other trajectories connecting these points can be used as well.} Let us show a reverse transition for $d = 8.757R$: starting at a REP $\omega_{\text{EP}} = (2.982 - 0.7352)c/R$ considered in Sec.~\ref{sec:rep}, we smoothly tune $\omega_{\text{EP}}$ to $3.227c/R$. This exceptional point corresponds to \PT-symmetric poles $\omega_{1,2} = (3.227 \pm 0.0002i)c/R$.  Therefore, both kinds of EP emerge due to the same physical mechanism.  }

A similar effect, that is, an exceptional point emerging from the interaction of a dimer with the surrounding space, has been reported in elastodynamics~\cite{dominguezrocha2020environmentally}. However, in that study the environment was characterised by a discrete spectrum of vibrations a finite substrate, rather than a continuum.

  While the dimers considered in this paper are difficult to realize in an experiment, we anticipate that this system shows the physics of the exceptional points occurring without gain and loss. This understanding may prove instructive to design exceptional points in experimentally viable systems, such as dimers of silicon nanodisks~\cite{valero2021exceptional}.

\acknowledgements
This work was supported by the Russian Science Foundation (project \textnumero 21-79-10190).

\end{document}